\documentclass[12pt,aps,noshowpacs]{revtex4}%
\usepackage{amsmath}
\usepackage{amssymb}
\usepackage{amsfonts}
\usepackage{graphicx}%
\providecommand{\U}[1]{\protect\rule{.1in}{.1in}}
\begin{document}
\title{Generalized Inheritance}
\author{J.P Krisch and E.N. Glass}
\affiliation{Physics Department, University of Michigan}
\date{August 18, 2015}

\begin{abstract}
Generalized inheritance is used with the almost-conformal Killing

equation. Examples are the flat FRW, Kasner, and deSitter metrics. The

volume changes in FRW while transitioning to a stiff fluid are

discussed. An inheritance current is implicit in the generalized

condition.\newline\ \newline Keywords: inheritance, ACKV, FRW, deSitter

\end{abstract}
\maketitle

\section{Introduction}

Solutions and models in general relativity are increasingly being generalized
to include more physically realistic matter content and symmetry descriptions.
The symmetries associated with Killing vectors and conformal Killing vectors
are defined though the Lie derivative of the metric
\begin{equation}
\mathcal{L}_{\xi}g_{ab}=\xi_{a;b}+\xi_{b;a}=\kappa(\nabla\cdot\xi)g_{ab}
\label{Lie-1}%
\end{equation}
with zero divergence defining the Killing vector (KV) and $\kappa=1/2$ the
conformal Killing vector (CKV). Other types of generators are classified by
their divergence behavior:
\begin{subequations}
\begin{align}
(\nabla\cdot\xi),_{a}  &  =0,\text{ homothetic vector}\\
(\nabla\cdot\xi),_{ab}  &  =0,\text{ }(\nabla\cdot\xi),_{a}\neq0,\text{special
conformal vector}\\
(\nabla\cdot\xi),_{ab}  &  \neq0,\text{ proper conformal vector}%
\end{align}
The metric symmetries do not automatically extend to other curvature functions
\cite{KY57,ST91,YYT+99,Cam02} or to matter sources in the spacetime
\cite{HJL+84, CT90a,CT90b,CT90c,TM90,BY02,Sha06,MS14,Smo15}.\ For a fluid with
unit flow vector $\hat{U}^{a}$ and a CKV satisfying Eq.(\ref{Lie-1}), Maartens
et al \cite{MMT86,MM86} established the kinematic result
\end{subequations}
\begin{equation}
\mathcal{L}_{\xi}\hat{U}^{a}=-\frac{\kappa}{2}(\nabla\cdot\xi)\hat{U}%
^{a}+V^{a} \label{Lie-U-eqn}%
\end{equation}
where $V^{a}\hat{U}_{a}=0$ and $\hat{U}_{a}\hat{U}^{a}=-1$.\ It is a kinematic
constraint which compares velocity parameters in generalized matter content
across general relativity solution sets \cite{MM87, ST91,YY97}. Coley and
Tupper \cite{CT90a,CT90b,AAC91} used the $V^{a}=0$ case to define
\textit{inheritance} as the condition which conformally maps flow lines onto
flow lines, i.e.
\begin{equation}
\mathcal{L}_{\xi}\hat{U}^{a}=-\frac{\kappa}{2}(\nabla\cdot\xi)\hat{U}^{a}
\label{Lie2-U-eqn}%
\end{equation}

Several generalizations of the conformal Killing equation have been suggested.
Matzner \cite{MAT68} suggested
\begin{equation}
\nabla_{b}[\xi^{(a;b)}]\sim const\cdot\xi^{a}. \label{ckv-eqn}%
\end{equation}
Later work \cite{Tau78,BCP05} considered generalizations of the $\kappa=0$
Yano-Bochner equation \cite{YB53,KY58}, defining an almost CKV (ACKV). \
\begin{equation}
\nabla_{a}[\bar{\xi}^{a;b}+\bar{\xi}^{b;a}-\kappa(\nabla\cdot\bar{\xi}%
)g^{ab}]=0 \label{div-CKV-eqn}%
\end{equation}
with $\kappa$ generalizing the CKV divergence contribution ($\xi^{a}$ is a
CKV, $\bar{\xi}^{a}$ is an ACKV).

Using the Riemann tensor to write the double derivatives, the ACKV equation
can also be written as%
\begin{equation}
\nabla_{b}\nabla^{b}\bar{\xi}^{a}+R_{\ b}^{a}\bar{\xi}^{b}+(1-\kappa
)\nabla^{a}(\nabla\cdot\bar{\xi})=0 \label{ACKV-eqn}%
\end{equation}

The "almost" idea has been used in several contexts.\ A common use is to
describe an "almost symmetry" as one that asymptotically becomes a metric
symmetry \cite{Tau78}. Another is to consider symmetry generators on surfaces
that are distorted from spherical symmetry \cite{ABL02, CW07}, or generators
in perturbed manifolds with an exact symmetry \cite{CB08,BW14}.\ Just as the
conformal Killing equation was generalized with a divergence, the
\textit{inheritance} condition can be extended with a divergence.\ 

Here we suggest a generalization of the \textit{inheritance} condition by
taking a divergence of the mapping equation for the four-velocity%
\begin{equation}
\nabla_{a}[\mathcal{L}_{\bar{\xi}}\hat{U}^{a}+\frac{\kappa}{2}\hat{U}%
^{a}(\nabla\cdot\bar{\xi})]=0. \label{div-four-vel}%
\end{equation}
The Lie derivative has been scaled to match the scaling in the ACKV equation,
Eq.(\ref{div-CKV-eqn}).\ \textit{Inheritance }is implicit in the conformal
Killing equation for proper, timelike CKV's \cite{AAC91}%
.\ Coupling\ "almost-inheritance" with "almost-symmetry", examines the role of
inheritance for vectors obeying the almost-conformal-Killing equation and
allows for scalings beyond the conformal $\kappa=1/2$. We discuss the effects
which follow from a generalized inheritance condition for a timelike vector,
$\xi^{a}=\chi\hat{U}^{a}$. For this case, one difference will be in the value
of the scaling parameter $\kappa$. Another is the use of $\kappa$ as a
modeling parameter for homothetic vectors. For $\nabla\cdot\xi=const$,
$\kappa$ does not appear in the ACKV equation,\ Eq.(\ref{div-CKV-eqn}), but
does remain in the almost-inheritance condition. This allows some tracking of
symmetry transitions. In the next section we consider the ACKV parameters. The
ACKV equation and almost inheritance are discussed in the third part of the
paper with some metric examples.

\section{Almost Inheritance}

\subsection*{Conformal Parameters}

A timelike CKV is $\xi^{a}=\chi\hat{U}^{a}\ $with parameters $\xi^{a}\xi
_{a}=-\chi^{2}$, $\kappa=1/2$, and velocity parameters of the CKV observer.
One of the differences between a KV, a CKV, and an ACKV is the scaling along
the flow of the vector trajectory.\ For Killing and conformal Killing vectors,
Eq.(\ref{Lie-1}) projected by $\hat{U}^{a}$ and $\hat{U}^{b} $ relates the
divergence to the change in the norm \
\begin{equation}
\kappa(\nabla\cdot\xi)=2\dot{\chi} \label{xi-dot}%
\end{equation}
KVs with zero divergence have constant norm along their trajectory.\ The CKV
norm change is related to the divergence with $\kappa=1/2.$ For KVs and\ CKVs,
the expansion of the observer velocity is also related to $\dot{\chi}.$ Using
Eq.(\ref{xi-dot}), the expansion $\Theta:=\nabla_{a}\hat{U}^{a}$ is
\begin{equation}
\Theta=3(\dot{\chi}/\chi). \label{theta-eqn}%
\end{equation}
A direct velocity projection is not possible with the ACKV equation and there
is no simple relation between the ACKV observer expansion and derivatives of
$\chi$. Next we consider possible solutions to the almost inheritance
condition and possible relations between $\Theta$, $\chi$, and its derivatives
for ACKV $\bar{\xi}^{a}$.

\subsection*{General Parameters}

For a timelike ACKV, $\bar{\xi}^{a}=\chi\hat{U}^{a}$, the generalized
inheritance condition is%
\begin{equation}
\nabla_{a}\left[  \hat{U}^{a}[\frac{(\kappa-2)}{2}\dot{\chi}+\frac{\kappa}%
{2}\chi\Theta]\right]  =0 \label{gen-inherit-eqn}%
\end{equation}
or, in expanded form.%
\begin{equation}
(\frac{\kappa-2}{2})\ddot{\chi}+\Theta(\kappa-1)\dot{\chi}+\frac{\kappa}%
{2}(\dot{\Theta}+\Theta^{2})\chi=0 \label{almost-inherit}%
\end{equation}
with $\dot{\chi}=\chi_{,a}\hat{U}^{a}$. There are several solutions.\ An
obvious one, which follows from Eq.(\ref{gen-inherit-eqn}) by inspection, is
\[
\Theta=(\frac{2-\kappa}{\kappa})\frac{\dot{\chi}}{\chi}%
\]
a generalization of the conformal Eq.(\ref{theta-eqn}). A further
generalization of this choice is \
\begin{equation}
\Theta=\Theta_{1}(\frac{\dot{\chi}}{\chi})
\end{equation}
With these choices for $\Theta$, the generalized inheritance condition can be
written as a product.
\[
(\ddot{\chi}+\Theta_{1}\frac{\dot{\chi}^{2}}{\chi})\left[  (\frac{\kappa-2}%
{2})+\frac{\kappa}{2}\Theta_{1}\right]  =0
\]
There are two solutions.
\begin{align}
(1)\text{ \ }\Theta_{1}  &  =\frac{2-\kappa}{\kappa}\label{soln-1}\\
(2)\text{ \ }\Theta_{1}  &  =-(\frac{\ddot{\chi}}{\dot{\chi}})(\frac{\chi
}{\dot{\chi}}) \label{soln-2}%
\end{align}
with the second solution implying the relation $\Theta=-\ddot{\chi}/\dot{\chi
}.$ When the ACKV has constant divergence, the modeling relation
\begin{equation}
\nabla\cdot\bar{\xi}=\dot{\chi}+\chi\Theta=\dot{\chi}(1+\Theta_{1})=\delta
\end{equation}
implies $\dot{\chi}=const$ and almost inheritance eliminates solution (2) used
by itself. When both solutions are simultaneously true, the almost inheritance
condition implies no expansion.
\begin{equation}
(\frac{\dot{\chi}^{2}}{\chi^{2}})\Theta_{1}=0
\end{equation}
There are other solution limits when used with the complete general
inheritance condition.\ For example, the second solution implies%
\[
\frac{\kappa}{2}(-\frac{\ddot{\chi}}{\chi}+\dot{\Theta}+\Theta^{2})=0
\]
For metrics with $\dot{\Theta}+\Theta^{2}=0,\ $either $\kappa=0$ or
$\ddot{\chi}=0,$ with $\dot{\chi}=const$. For these metrics\
\begin{equation}
\Theta(\kappa-1)\dot{\chi}+(\frac{\kappa-2}{2})\ddot{\chi}=0
\end{equation}
Observers with a non-zero expansion and constant $\dot{\chi}$ would have
$\kappa=1.$ There are several interesting solutions that satisfy this
condition and will be discussed in the next section.

\section{ACKV Examples}

The examples we consider are flat FRW, Kasner, and deSitter metrics.\ For
geodesic and irrotational metrics the ACKV equation, Eq.(\ref{ACKV-eqn}), is%
\begin{equation}
\chi_{b;a}\hat{U}^{a}+\chi_{b}\Theta-\hat{U}_{b}\nabla_{a}\chi^{a}-\chi
^{a}\hat{U}_{b;a}-(\kappa-2)\nabla_{b}(\nabla\cdot\bar{\xi})=-2R_{ba}\bar{\xi
}^{a}%
\end{equation}
Projecting with $\hat{U}^{b}$ and spatial vector $S^{b}$, the equations are%
\begin{align}
\chi_{b;a}\hat{U}^{a}\hat{U}^{b}+\dot{\chi}\Theta+\nabla_{a}\chi^{a}%
-(\kappa-2)\hat{U}^{b}\nabla_{b}(\nabla\cdot\bar{\xi})  &  =-2\chi R_{ba}%
\hat{U}^{a}\hat{U}^{b}\\
\chi_{b;a}^{\ }\hat{U}^{a}S^{b}+\chi,_{b}\Theta S^{b}-\chi^{a}S^{b}\hat
{U}_{b;a}-(\kappa-2)S^{b}\nabla_{b}(\nabla\cdot\bar{\xi})  &  =-2\chi
R_{ba}S^{b}\hat{U}^{a}%
\end{align}

\subsection*{FRW}

Flat FRW is a simple example of solution (2) metric. \
\begin{equation}
ds^{2}=-dt^{2}+A^{2}(t)(dx^{2}+dy^{2}+dz^{2})
\end{equation}
with $\hat{U}^{a}=[1,0,0,0]$ and $\Theta=3\dot{A}/A$.\ The timelike ACKV
equation is
\begin{align}
\ddot{\chi}+\dot{\chi}\Theta+\nabla_{a}\chi^{a}-(\kappa-2)\hat{U}^{b}%
\nabla_{b}(\nabla\cdot\bar{\xi})=-2\chi R^{ba}\hat{U}_{a}\hat{U}_{b}  & \\
\frac{(\kappa-2)}{\chi}\hat{U}^{t}\nabla_{t}[\chi_{,t}+3\chi\frac{\dot{A}}%
{A}]=-6\frac{\ddot{A}}{A}  &
\end{align}
Let $A=A_{0}t^{n},$ with for example, $n=2/3$ FRW dust and $n=1/2$. This is
the Tolman radiation solution \cite{SKM+03}.\ The second almost inheritance
solution, Eq.(\ref{almost-inherit}), for the comoving expansion and $\chi
=\chi(t)$ is
\begin{align}
\chi &  =[(\Theta_{1}+1)(c_{1}t+c_{2})]^{\frac{1}{\Theta_{1}+1}}%
\label{chi-soln}\\
\Theta &  =\frac{\Theta_{1}}{1+\Theta_{1}}\left(  \frac{c_{1}}{c_{1}t+c_{2}%
}\right)  \label{theta-soln}%
\end{align}
where $(c_{1},c_{2})$ are assumed constant. Taking $c_{2}=0$, and comparing to
the FRW expansion $\Theta=3(\dot{A}/A),$ $\Theta_{1}$ is determined from the
known solutions.
\begin{equation}
\Theta_{1}=\frac{3n}{1-3n}%
\end{equation}
Using these the ACKV equation relates $\kappa$, $\Theta_{1}$ and the FRW
power, $n$. \
\begin{align}
-(\kappa-2)(\frac{1}{\Theta_{1}+1}-1)(\frac{1}{\Theta_{1}+1}+3n)  &
=6n(n-1)\\
\kappa &  =2n\nonumber
\end{align}
The parameter set is determined in general by the power law behavior of the
metric, $\Theta_{1}=3n/(1-3n),$ $\kappa=2n$. For dust, the values are
$\Theta_{1}=-2,$ $\kappa=4/3$ and for radiation $\Theta_{1}=-3,$ $\kappa=1.$
The spatial ACKV equation is identically zero. The stress-energy components
for this FRW metric are
\begin{subequations}
\begin{align}
8\pi\rho &  =3n^{2}/t^{2}\\
8\pi P  &  =n(2-3n)/t^{2}%
\end{align}
$1/3\leq n\leq2/3$ is required for physical stress and acoustic speed $\leq
1$.\ The stress-energy is well behaved for the stiff fluid value, $n=1/3$, but
$\Theta_{1}$ is not. The stiff fluid with $n=1/3$ has expansion $\Theta=1/t$
and satisfies the relation $\dot{\Theta}+\Theta^{2}=0.$ This is the special
solution discussed in Section 2 where the ACKV has a constant divergence
together with the stiff fluid case $\Theta_{1}=1,$ $\kappa=1.$ The
discontinuity in $\Theta_{1}$ is related to the behavior of the\ FRW spatial
volume, $V$. \
\end{subequations}
\[
dV=t^{3n}dV_{3}%
\]
with $dV_{3}=dxdydz.$ The way the volume changes along the FRW flow is
\[
\dot{V}=3nt^{3n-1}dV_{3}%
\]
and for the stiff perfect fluid, $\dot{V}=const$. The transition from a
non-stiff to a stiff equation of state reflects a change in the volume
structure of the fluid. This is like the transition from a body centered cubic
structure to a face centered cubic structure in a heated iron wire.\ Here the
symmetry change is the appearance of the homothetic ACKV \cite{CM1975,CM1976}
for the stiff fluid and the transition from solution (2) to solution (1) with
the associated discontinuities in the parameter sets. \ 

\subsection*{Kasner}

Kasner vacuum, with four-velocity $\hat{U}^{a}=[1,0,0,0],$ is an example of a
solution (1) metric. \ \
\[
ds^{2}=-dt^{2}+t^{2p_{1}}dx^{2}+t^{2p_{2}}dy^{2}+t^{2p_{3}}dz^{2}%
\]
with $p_{1}+p_{2}+p_{3}=1$ and $p_{1}^{2}+p_{2}^{2}+p_{3}^{2}=1$. The metric
is geodesic and irrotational with expansion $\Theta=1/t\ $such that
$\dot{\Theta}+\Theta^{2}=0.$ The timelike ACKV equation is
\begin{subequations}
\begin{align}
\chi_{t,t}+\chi_{t}/t-\chi_{tt}-\chi_{t}/t  &  =0\\
(\kappa-2)\hat{U}^{b}\nabla_{b}(\nabla\cdot\bar{\xi})  &  =0
\end{align}
and requires that either $\bar{\xi}^{a}$ has a constant divergence or
$\kappa=2$ with the divergence not determined.\ If the divergence is not
constant, the first almost inheritance condition enters and, with $\kappa=2, $
requires zero expansion. The constant divergence determines $\chi$
\end{subequations}
\begin{align}
\dot{\chi}+\frac{\chi}{t}  &  =\delta\text{ \ or \ }\frac{(t\chi),_{t}}%
{t}=\delta\text{, \ thus}\nonumber\\
\chi &  =\frac{\delta t}{2}+\frac{c_{1}}{t}%
\end{align}
The almost inheritance condition determines $\kappa$
\begin{equation}
\frac{2c_{1}}{t^{3}}+(\frac{\delta}{2}-\frac{c_{1}}{t^{2}}-\frac{\kappa}%
{2}\delta)\frac{1}{t}=0,\text{ \ }c_{1}=0,\text{ \ }\kappa=1
\end{equation}
The spatial ACKV equation is an identity. Kasner has a set of homothetic
vectors \cite{TPK15}$\ $with the ACKV form part of the known homothetic CKV.
The relation between the expansion and $\chi$ is $\Theta=\dot{\chi}/\chi$ with
$\Theta_{1}=1.$ The Kasner parameter set is $\Theta_{1}=1,$ $\kappa=1.$ There
is no phase behavior in\ Kasner since, even with anisotropic volume expansion,
the volume has the same form over the parameter range\ $dV=t^{p1+p2+p3}%
dV_{3}=tdV_{3}.$

\subsection*{deSitter}

The third example is the deSitter form of the FRW metric with $A(t)=A_{0}%
e^{\alpha t}.$\ The expansion is constant, $\Theta=3\alpha$ and $\chi$ follows
as%
\[
\chi=\chi_{0}\exp(3\alpha t/\Theta_{1}).
\]
The almost inheritance condition is
\begin{subequations}
\begin{align}
(\frac{\kappa-2}{2})\frac{9\alpha^{2}}{\Theta_{1}^{2}}+3\alpha(\kappa
-1)\frac{3\alpha}{\Theta_{1}}+\frac{\kappa}{2}(9\alpha^{2})  &  =0\\
\kappa-2+2(\kappa-1)\Theta_{1}+\kappa\Theta_{1}^{2}  &  =0
\end{align}
with solutions $\Theta_{1}=(2-\kappa)/\kappa$ and $\Theta_{1}=-1.$ The ACKV
equation is
\end{subequations}
\begin{subequations}
\begin{align}
(2-\kappa)(\frac{3\alpha}{\Theta_{1}}+3\alpha)\frac{3\alpha}{\Theta_{1}}  &
=6\alpha^{2}\\
(2-\kappa)(\Theta_{1}+1)  &  =\frac{2}{3}\Theta_{1}^{2}%
\end{align}
The deSitter parameters are the original conformal set $\kappa=1/2,$
$\Theta_{1}=3.$

\section{Inheritance Current}

Currents can be associated with Killing structures. Ruiz, Palenzuela and Bona
\cite{RPB13} used a Killing vector with a stress-energy to define a conserved
current
\end{subequations}
\begin{equation}
J_{RPB}^{a}=T^{ab}\xi_{b}%
\end{equation}

The generalized inheritance condition, Eq.(\ref{div-four-vel}), was
constructed from the inheritance condition, Eq.(\ref{Lie2-U-eqn}). The zero
divergence can also be used to define a conserved inheritance current%
\begin{equation}
J^{a}=\mathcal{L}_{\bar{\xi}}\hat{U}^{a}+\frac{\kappa}{2}\hat{U}^{a}%
(\nabla\cdot\bar{\xi})
\end{equation}
For the timelike ACKV considered here, the inheritance current is
\begin{equation}
J^{a}=\left[  (\frac{\kappa-2+\kappa\Theta_{1}}{2})\dot{\chi}\right]  \hat
{U}^{a}=\left[  \Theta(\frac{\kappa-2+\kappa\Theta_{1}}{2\Theta_{1}}%
)\chi\right]  \hat{U}^{a}%
\end{equation}
and, with a dependence on expansion, describes the rate of change of a
geodesic cross section \cite{Poi04}. The current is zero for solution
(1).\ For non-stiff FRW, using the relation between parameters, $\kappa
=2\Theta_{1}/[3(1+\Theta_{1})],$ the inheritance current reflects the CKV
parameter structure and is also zero for $\Theta_{1}=\Theta.$
\begin{equation}
J^{a}=\left[  (\frac{\Theta_{1}-3}{3})\dot{\chi}\right]  \hat{U}^{a}%
\end{equation}

The FRW solution has volume transition behavior in moving to the stiff fluid.
The FRW inheritance current is non zero for the non-stiff fluid with its
changing cross section. \
\begin{equation}
J^{a}=\left[  \dot{\chi}\frac{4n-1}{1-3n}\right]  \hat{U}^{a},\text{ \ }n>1/3.
\end{equation}
and reflects the same volume discontinuity at $n=1/3$. \ 

Consider a small region of the FRW space interior to a hypersurface,\ $S^{a}$
with normal $\hat{U}^{a}$. For zero divergence, Gauss' theorem provides the
integral of the current over the hypersurface boundary%
\begin{equation}%
{\textstyle\oint}
J^{a}dS_{a}=0
\end{equation}
This implies that the integral is hypersurface independent.\ Substituting we
have%
\begin{equation}%
{\textstyle\oint}
\frac{1-4n}{1-3n}\dot{\chi}\ t^{3n}dV_{3}=0
\end{equation}
Using $\Theta=\Theta_{1}(\dot{\chi}/\chi)$ and the solutions,
Eqs.(\ref{chi-soln},\ref{theta-soln}), the integral can be written as
\begin{equation}%
{\textstyle\oint}
(1-4n)\frac{(1-3n)^{3n-1}c_{1}^{\frac{1}{\Theta_{1}+1}}}{t^{3n}}\ t^{3n}%
dV_{3}=0
\end{equation}
The current can be interpreted as maintaining the size of the base coordinate
3-volume. \ 

\section{Discussion}

We have suggested an almost inheritance condition to be used with the ACKV
equation.\ It is a two parameter, $[\theta_{1},\kappa]$, model \ generalizing
the conformal relation between the observer expansion and the ACKV norm with
parameters $[\theta=3,$ $\kappa=1/2]$. With this generalization, two solutions
for generalized inheritance were presented and FRW, Kasner and de Sitter were
used as examples.\ The three metrics were chosen because they span the class
of the almost-inheritance solutions and because they all appear frequently in
current models.\ Almost inheritance, applied to these spacetimes, can extend
the range of the many current applications.\ The FRW metric is frequently used
in general relativity extensions. \cite{Mir03, MS+10, RMC15}. The extended
parameter set allowed by the generalized inheritance condition extends the
range of perfect fluid spaces, linking the equation of state to the almost
conformal parameter. A new behavior is the possible phase transition in
transfer to a stiff fluid.\ The Kasner metric is used in models of the very
early universe \cite{KUP11,BM15} and is useful in considering anisotropic
volume expansion \cite{CAM11}. As a background metric it is used to explore
solutions of differential equations \cite{Sri89, KBK+90} and in developing
extended classifications \cite{TPK15}. The deSitter universe, with accelerated
expansion, is an important model of both our early \cite{SMO+09} and our
asymptotic universe \cite{Mal13}.\ Like the Kasner metric, it is used as a
background space to explore equation solutions\ \cite{AN05, BK05, CAN12} and
to support deSitter process descriptions \cite{MTW09, LPW13}.

By themselves the solutions presented here are useful in discussing divergence
behaviors for the ACKV. In the almost-inheritance model, Kasner has\ a
constant divergence ACKV with the solution (1) parameter set $[1,1]$. The
power law FRW example, $A(t)=A_{0}t^{n}$, used both solutions, with the
non-stiff perfect fluids having parameter set $[3n/(1-3n),2n]$. The stiff
fluid solution (1) has set $[1,1]$. deSitter has the original conformal parameters.\ 

The four-velocities in the examples used were all simple comoving identities.
Outgoing, marginally bound geodesic Schwarzschild \ \cite{Poi04} with
four-velocity
\[
\hat{U}^{a}=[(1-2m_{0}/r)^{-1},\sqrt{2m_{0}/r},0,0]
\]
has expansion $\Theta=(3/2)\sqrt{\ 2m_{0}/r^{3}}$ and also obeys the condition
$\dot{\Theta}+\Theta^{2}=0.$ Studying the relation between ACKVs and
generalized inheritance for velocities with more structure would be of
interest. \

\end{document}